# Hierarchical Entity Alignment for Attribute-Rich Event-Driven Graphs[*]

Elizabeth Hou[†‡]   Joanna Brown[†]   John Fisher[§]

**Abstract**

This paper addresses the problem of entity alignment in attribute-rich event-driven graphs. Unlike many other entity alignment problems, we are interested in aligning entities based on the similarity of their actions, i.e., entities that participate in similar events are more likely to be the same. We model the generative process of this problem as a Bayesian model and derive our proposed algorithm from the posterior predictive distribution. We apply our Hierarchical Entity AlignmenT (HEAT) algorithm to two datasets, one on publications and the other on financial transactions, derived from real data and provided to us by an external collaborator.

## 1  Introduction

The ability to represent information as graphs is appealing as it allows for an intuitive way to store highly complex, relational data. Thus, it is natural that graphs are extremely prevalent in various domains in the real world. Some of the most well studied graphs include social networks that depict relationships in various social media platforms, traffic networks that depict the sharing of messages, and Gaussian graphical models that depict conditional independence relationships for various sorts of data such as gene expression or brain imaging.

However, the information stored in these graphs is almost always noisy and incomplete due to noise in the observations themselves, errors in the data collection process, or an inability to observe latent factors of the data. Consequently the ability to align nodes or complete missing information about the nodes is crucial to having a more clear understanding of the relationships in the data. In this paper, we will focus specifically on the problem of alignment. However, because we propose a generative model, we can also easily apply our model to the problem of information completion, which we will touch upon at the end.

---
[*]DISTRIBUTION STATEMENT A. Approved for public release, Distribution Unlimited.

[†]STR, 600 West Cummings Park, Woburn, MA 01801

[‡]Corresponding Author: elizabeth.hou@str.us

[§]Computer Science & Artificial Intelligence Laboratory, Massachusetts Institute of Technology, 32 Vassar St, Cambridge MA 02139

**1.1  Outline** The rest of the paper is organized in the following way. Section 2 describes the problem motivation. Section 3 describes our proposed model, first by formulating the problem mathematically and listing our assumptions, then by defining our solution as a Bayesian model, and finally by explicitly deriving our proposed algorithm. Section 4 contains the experiments where we describe the dataset used and show the performance results of our proposed model. Finally, Section 5 is a discussion that concludes the paper.

## 2  Motivation

Due to the prevalence of graphical models, the general problem of alignment in graphs is ubiquitously acknowledged as an important field of research. However, solutions to this problem are context dependent due to the diverse set of use cases which result in different graphical representations.

One such use case is the Defense Advanced Research Projects Agency's (DARPA) Modeling Adversarial Activity (MAA) program, which is interested in detecting weapons of mass terror (WMT) threats through integrating multiple sources of transaction data. While the data used in such analysis readily lends itself to being modelled as graphs, a crucial element for generating insight from the graphs is accurate graph alignment and merging. Thus, one of the technical areas of the MAA program is dedicated to automated alignment of graph data sets into an integrated worldview.

As the MAA program is specifically interested in representing adversarial events in graphs, these graphs must be attribute-rich to effectively represent event-driven actions. The attribute-richness is necessary as the events are defined by the relationships between involved entities which are in turn defined by their attributes. However, defining events through entities gives rise to an alignment problem as entities with noisy attributes may inadvertently appear distinct. Alignment of richly attribute graphs is not well studied; thus we propose a novel solution, the Hierarchical Entity AlignmenT (HEAT) algorithm, which is constructed with a Bayesian generative model.

Necessary to understanding our proposed model, it is essential to understand how information surrounding events is being represented in a graph. We will first discuss storing event-driven relationships in a specific graph representation, which we call the "fact" graph. Then we discuss an equivalent representation that does not explicitly incorporate the attributes of the "fact" graph, but satisfies the conditions of a traditional probabilistic graphical model.

The information we want to store are relations to an event node, thus we create edges between event and entity nodes to store their relationship in the graph. Any edge between an event node and an entity node can be viewed as a fact that has occurred because they represent an action where the entity is participating in the corresponding event. For example, event node Pub75180 represents the publication of a paper where entity nodes Author97632 and Author97633 are co-authors of the paper. Thus we can view this as two separate facts regarding the publication event; "Pub75180" "has author" "Author97632", and "Pub75180" "has author" "Author97633".

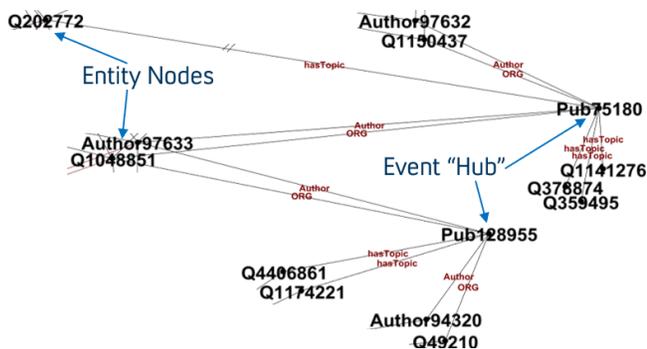

Figure 1: The "fact" graph representation where event "hub" nodes are the central components connecting the various entity nodes.

This framing is heavily influenced by knowledge base literature from classical artificial intelligence (AI) built upon the semantic triple, or Resource Description Framework (RDF) triple, which uses these "subject", "predicate", "object" expressions to codify a statement. Thus because the event nodes are central to the information we are representing, but are completely defined by their relations, they are always the "subjects", we denote them as event "hubs". Correspondingly, the entity nodes are always the "objects" of the fact triple, while the predicate relation is the connecting edge. Because the event nodes bind a set of entities to it through edges on a directed graph, but do not have any attributes themselves, we can also represent an event as a clique of its "object" nodes. Figure 2 shows the directed semantic graph above as an undirected clique of the entity nodes surrounding the event "hub" Pub75180.

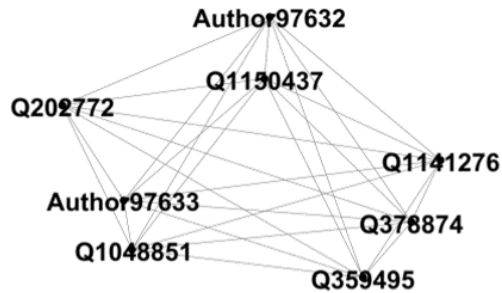

Figure 2: The clique representation where the events are implicitly represented as cliques of entity nodes.

The undirected clique representation has drawbacks, namely it does not explicitly represent events, but instead implicitly represents them through complete subgraphs, making it much harder to search the graph for specific events. However, because it more closely aligns with a traditional probabilistic graph, it is significantly easier to represent mathematically and formalize properties on, such as independence and neighborhood.

Unlike an RDF graph, one built directly from triples, both the "fact" graph in Figure 1 and the probabilistic graph in Figure 2 encode only relations between entities and events in their structure. However, because the "fact" graph is a rich attribute graph, we can instead encode other ontological definitions as various attributes on the nodes and edges of the graph. For example, if we are interested in encoding type or information for "Author97633", we can add triple-like attributes on the node as "Author97633" "is" "Human" or "Author97633" "has name" "Bob". The distinction between these triples and corresponding triples in an RDF graph is that, as attributes, they are not part of the structure of the "fact" graph.

In this paper we mainly focus on the probabilistic graph representation, but occasionally refer to the "fact" graph for intuitive understanding as this representation provides the motivation behind our proposed solution.

**2.1 Related Work** While research in the alignment of event-driven graphs is not very extensive, the alignment problem has been well studied for other types of graphs. Our formulation of a richly attributed event-driven graph shares some characteristics with the RDF based knowledge graphs prominently used in classical AI. Thus we will briefly review some works in this community that focus on entity alignment for their graph representations and discuss how they differ from ours. For a more comprehensive review of the knowledge graph AI literature, we refer to this recent survey paper [5].

The earliest works looking at the entity alignment problem focused on variations of string similarity [9, 15, 18, 13, 10, 12] where a variety of similarity functions were proposed. In more recent years, embedding based methods [1, 19, 6, 4, 16, 17, 22, 2, 21] have become more popular where the similarity is measured between the embeddings of a pair of entities instead of the entities themselves.

However, the information encoded in these embeddings is ontological information, not actions that an entity participates in. Because these methods are performing entity alignment on semantic graphs, e.g. RDF, they essentially are assuming entities are more similar if they are defined similarly according to their ontology. In contrast, we consider entities to be more similar if they perform similar actions. For example, the experiments in [16] are based on DBpedia and Wikidata graphs where the entities are connected to each other by statements such as "instance of" and "coordinate location". These types of graphs are incompatible with our use case where we are interested in actions such as authors publishing papers or people sending money to each other.

## 3 Model

In this section, we begin by defining the problem mathematically and enumerating the necessary assumptions to make estimation feasible. Then we describe our proposed generative model solution and link its formulation to the original motivating "fact" graph representation. Finally we explicitly derive our Hierarchical Entity AlignmenT (HEAT) algorithm and provide pseudocode for it.

**3.1 Problem Formulation** Let $\mathcal{G}'$ represent a probabilistic graph, such as the one shown in Figure 2 where the nodes correspond to entity nodes of the "fact" graph. We assume that the nodes in $\mathcal{G}'$ are statistical parameters that govern the generation of observable samples. Let $\mathcal{G}$ consist of noisy observations of this graph where multiple observed nodes may actually be variations of a single node. Thus the nodes in $\mathcal{G}$ are ambiguous because we cannot differentiate if they are the same node or not i.e. we cannot observe their true latent node ids. We assume the nodes in $\mathcal{G}$ are samples from a categorical distribution where each category corresponds to a node in $\mathcal{G}'$ or a new node altogether.

Let $n_i$ be a random variable corresponding to a node in $\mathcal{G}$ and let $n_k$ be a node in $\mathcal{G}'$. Then, by estimating the parameters of the categorical distribution we can calculate the probability that an ambiguous node in $\mathcal{G}$ is a noisy variation of a node in $\mathcal{G}'$, i.e. $n_i = n_k$. By making alignments between nodes $n_i$ and $n_k$ when there is high probability that they are equal and declaring a new node with there is not, we can merge the two graphs together. So for every node $n_i$ in $\mathcal{G}$, we denote its marginal conditional distribution $P(n_i|\mathcal{G}_{-i}, \mathcal{G}')$ to be from the categorical family.

However, the graphs $\mathcal{G}'$ and $\mathcal{G}$ lie in an extremely high dimensional space (each node can be viewed as a dimension) making estimation of the parameters of $P(n_i|\mathcal{G}_{-i}, \mathcal{G}')$ infeasible. Thus we make two critical assumptions on the low dimensionality and sparsity of the graph.

Definition 3.1.

*(A1) Each graph satisfies Markov properties [8] such that the number of influential nodes is small*

*(A2) The number of potential alignments with non-zero probabilities is sparse*

The distribution of $n_i$ depends on the other nodes in two places, implicitly through the other nodes' influences on $n_i$ and explicitly in the categories themselves. The above two assumptions reduce the number of dependent nodes in the two places respectively.

Because the probabilistic graph is constructed via a transformation of the "fact" graph, which is built from RDF triples, we inherit specific properties that are more obvious in the "fact" representation. Because entity nodes are only connected to the event nodes, an entity node is independent or unconnected with the rest of the graph when conditioned on its neighboring event nodes. And, although we cannot directly observed the event nodes, an event is fully described by the entity nodes participating in the event. Thus by assuming that an entity can be fully described by the events it participates in, which in part is described by the other entities in the event, the "fact" graph induces a probabilistic graph that is a Markov random field. By the graph separation property [8], we define the Markov blanket of node $n_i$ as

$$(3.1) \qquad \mathcal{B}_i = \{n_j | (n_i, n_j) \text{ share an edge}\}$$

where any node $n_j$ that is a neighbor of $n_i$ participates in the same events in the "fact" representation. Similarly, we denote $\mathcal{B}'_k$ to be a Markov blanket that separates $n_k$ from the rest of its graph $\mathcal{G}'$.

However, while we can use $\mathcal{B}_i$ so that the marginal conditional distribution $P(n_i|\mathcal{B}_i, \mathcal{G}')$ is not dependent on $\mathcal{G}$, the categories of the distribution represent the nodes of $\mathcal{G}'$. If we assume the categorical distribution is sparse, most nodes in $\mathcal{G}'$ have a probability of generating $n_i$ that

approaches 0, then the number of categories that are actually relevant is small. We define the set of nodes that correspond to these relevant categories to be an alignment set,

(3.2) $$\mathcal{K}_i = \{n_k | n_k \in C(n_i)\}$$

where $C(n_i)$ constrains the nodes in $\mathcal{K}_i$ to be only those that satisfy certain constraints with respect to $n_i$. Examples of such constraints will be given in the algorithms subsection.

Thus if assumptions (A1) and (A2) hold, the following distributions are equivalent

$$P(n_i|\mathcal{G}_{-i}, \mathcal{G}') = P(n_i|\mathcal{K}_i, \mathcal{B}_i, \cup_{k=1}^{K}\mathcal{B}'_k) = \prod_{k=1}^{K} p_k^{[n_i=n_k]}$$

where $[\cdot]$ is an Iverson bracket or indicator function and $\sum_{k=1}^{K} p_k = 1$ for $K = |\mathcal{K}_i|$.

**3.2 Proposed Solution** We propose a generative model that assumes a series of statistical distributions govern the generation of a graph. This gives us a principled way to estimate the probabilities of alignment between the nodes from $\mathcal{G}$ to nodes in $\mathcal{G}'$. We take a Bayesian approach as this gives us the ability to incorporate past information about the probabilities of alignment into our model. In particular, a Bayesian approach allows for the inclusion of the possibility that $n_i$ is a new node that has never been observed before.

Because we model the likelihood functions as categorical distributions, we assume Dirichlet priors in order to have conjugacy with the likelihood. This allows for a closed form expression of the posterior distribution and efficient direct estimation of its parameters. With different hyperparameters $\boldsymbol{\alpha}$ for the Dirichlet prior, we can represent different shapes for the prior distribution. For example, if all of its concentration parameters equal 1, the Dirichlet prior is equivalent to a uniform distribution over the categories. Whereas if all its concentration parameters equal 0 the prior is non-informative, i.e., it places all weight on the likelihood observations.

In the previous subsection, we showed how a Markov blanket $\mathcal{B}'_k$ prevents a node $n_k$ from being influenced by the rest of its graph $\mathcal{G}'$. Thus because all information about a node is in its neighbors, we explicitly define the form of a node's likelihood function using its neighbors. Again, we refer back to the "fact" representation for a more intuitive explanation of the function. Let $n_k$ participate in events $\mathcal{E}$, then $\mathcal{B}'_k$ are the other entity nodes that also participate in $\mathcal{E}$ by the definition in (3.1). Because we treat each RDF triple (event hub connected to entity node in the "fact" graph) as a separate fact, then if there are T facts associated with the entity nodes $\mathcal{B}'_k$ that are connected to the event hubs $\mathcal{E}$, we have T independent samples. Each of these T samples capture information about the influences on $n_k$, so we can define equivalence of the traditional Iverson bracket $[\cdot]$ of the categorical distribution as

$$[n_i = n_k] = [A(n_j) = A(n^t)]$$

where $n_j$ a node in $\mathcal{B}_i$ and $n^t = \mathcal{B}'^t_k$ is a neighbor of $n_k$ associated with one of the T facts. The function $A(\cdot)$ extracts the attributes of a node. This equality implies an equivalence between the nodes being equal and an attribute occurring in both neighborhoods. Like the constraint function first defined in (3.2), examples of attribute functions will be given in the algorithms subsection.

So far, we have treated all entity nodes as identically distributed. However, the "fact" graph is an attribute-rich graph and we need to encode this attribute information into the probabilistic graph itself. In particular, for neighborhood nodes, some may have more influence on $n_i$ than others. Thus, we extend our model to better reflect the "fact" graph by accounting for two factors: the rarity of an attribute and the importance of certain types of nodes.

Let $A(\mathcal{G}')$ be the attributes of all facts in $\mathcal{G}'$, then

(3.3) $$w_j = \frac{1}{\sum_{a \in A(\mathcal{G}')}[A(n_j) = a]}$$

is proportional to the number of times an attribute is observed in $\mathcal{G}'$, i.e., it is a weight measuring the rarity of an attribute $A(n_j)$ associated with a neighbor of $n_i$. We want to weight the probability of alignment between $n_i$ and $n_k$ so that observations with more common attributes are less important. In similar fashion to a generalized linear model (glm), we can model the categorical distribution to be a linear model in its parameters with explicitly defined coefficients $W_j = p_k^{w_j - 1}$ that are decreasing functions of the rarity of an attribute defined in (3.3). Thus we can decompose the linear model as

$$(W_j p_k)^{[n_i = n_k]} = (p_k^{w_j - 1} p_k)^{[n_i = n_k]} = p_k^{w_j [A(n_j) = A(\mathcal{B}'^t_k)]}$$

where the coefficient $W_j$ corresponds to a specific node $n_j$ that is a neighbor of $n_i$.

While $w_j$ essentially puts lower weight on facts involving more common attributes $A(n_j)$, we also need to incorporate the ability to differentiate between nodes that take on different "types" in the "fact" graph. Specifically certain types of neighborhood nodes may be so important that if they are missing, the probability of alignment goes to 0. Thus, we denote nodes of these types, which are a subset of the nodes in $\mathcal{B}_i$, as indicator variables. Let these indicator variables $\mathcal{I}_i$ be defined to have parameters (like in a mixture model) as

$$(3.4) \quad \iota_k = \max_{a \in A(\mathcal{I}_i)} \frac{\sum_{t=1}^{T}[a = A(\mathcal{B}'^t_k)]}{|\mathcal{E}|}$$

so that at $\iota_k$ is nonzero if at least one attribute of these indicator variables also exists in the neighbors of $n_k$.

Thus, if the prior distribution of the probabilities of alignment is $\mathbf{p} \sim \text{Dir}(K, \boldsymbol{\alpha})$ and the joint likelihood function of $n_i$ is a categorical distribution defined as

$$P(n_i | \mathcal{K}_i, \mathcal{B}_i, \cup_{k=1}^{K} \mathcal{B}'_k) = \prod_{k=1}^{K} p_k^{c_k}$$

then the posterior distribution of the probabilities of alignment is also a Dirichlet distribution $\mathbf{p} \sim \text{Dir}(K, \mathbf{c} + \boldsymbol{\alpha})$ where $\mathbf{c}$ is a vector of the number of occurrences or counts for each of the K categories. The counts $c_k$ are derived from the likelihood function as

$$(3.5) \quad c_k = \iota_k \sum_{n_j \in \mathcal{B}_i} \sum_{t=1}^{T} w_j [A(n_j) = A(\mathcal{B}'^t_k)]$$

The posterior predictive distribution that node $n_i$ is equivalent to any node in its alignment set $\mathcal{K}_i$ is a Dirichlet-categorical distribution defined as

$$(3.6) \quad P(n_i = n_k | n_k, \mathcal{B}_i, \mathcal{B}'_k, \boldsymbol{\alpha}) = \frac{c_k + \alpha_k}{\sum_{k=1}^{K} c_k + \alpha_k}$$

where $c_k$ is defined in (3.5), $\alpha_k$ is the prior parameter when observing node $n_k$, and T is the number of "facts" associated with the events that $n_i$ participates in. Intuitively, we can interpret this as the probability of alignment between $n_i$ and $n_k$ is proportional to the number of the observations of overlap between the attributes of the two's neighbors. Note, the posterior predictive probability of the Dirichlet-categorical distribution is equivalent to the expected value of the posterior distribution of $\mathbf{p}$ [7], which is why we have been interchangeably using estimating $\mathbf{p}$ with estimating the probabilities of alignment.

**3.3 Algorithm** In the previous subsections, we described our assumptions about a generative model and the estimation of its parameters. In this subsection, we use these theoretical underpinnings to explicitly derive our proposed algorithm.

The goal of the HEAT algorithm is to perform entity alignment between multiple graphs in order to form one unified graph. To simplify the notation, we will explicitly derive this algorithm for the alignment of one graph to another; however, we can consider the case of aligning entities within a graph to be one where the two graphs are the same, and the case of more than two graphs to be one of multiple pairwise alignments. Thus we will begin with deriving the central component of the HEAT algorithm, which performs entity alignment.

The Entity AlignmenT (EAT) algorithm takes in two graphs $\mathcal{G}$ and $\mathcal{G}'$ and outputs a list of predictions of the expected probabilities for each of the nodes in $\mathcal{G}$ to be aligned to the nodes in $\mathcal{G}'$. As the EAT algorithm is derived from a Bayesian model, it requires defining the prior distribution, which for a Dirichlet prior, can be fully defined by its parameters $\boldsymbol{\alpha}$. Because of the dimensionality reduction assumptions in (A1) and (A2), it also requires defining the lower dimensional space with constraints $C(\cdot)$, which forms the support of the likelihood functions. Finally, in order to preserve relevant attribute information from the "fact" graph, we must also define an attribute mapping function $A(\cdot)$. By extracting these data specific definitions to be inputs, the EAT algorithm is dataset independent and can be applied to any attribute-rich graph.

The choice in prior parameters $\boldsymbol{\alpha}$ can be dataset specific. If a user has no knowledge about either of the graphs, then the symmetric Dirichlet distribution $\text{Dir}(1, \ldots, 1)$ will uniformly distribute the prior belief among all probabilities of alignment. However, if a user believes that certain nodes in $\mathcal{G}$, e.g. $n_i$, are more likely to be aligned with certain nodes in $\mathcal{G}'$, e.g., $n_k$, then the user can give a higher value to the parameter for the prior distribution of $n_i$ that corresponds to the category associated with $n_k$ to reflect this belief, e.g., $\text{Dir}(1, \ldots, 5, \ldots, 1)$. Or, if a user does not have any prior beliefs and wants all weight to be focused on the observed neighbors, they can choose an non-informative prior $\text{Dir}(0, \ldots, 0)$.

The attribute information used for the EAT algorithm is specific to each dataset. Obviously, a dataset composed of dog interactions at parks will contain different attributes than a dataset composed of pop music concerts attend by teenagers. However, the importance and use of certain attributes can also depend on the user and their goals. If a user is interested in aligning dogs in graphs representing the attendance of dogs at specific parks, the user may believe that the dog breed and collar color are important attributes for their goal. Whereas if a user is interested in aligning dogs' owners at these parks, then the name, phone number, and address of the owner may be more significant attributes. Additionally name, cell phone number, and address are different "types" of entity nodes in that if they were in an ontology, they would point to different subclasses of entity. A user can also designate certain types, e.g.

phone numbers, as being indicative of being the same by including them in the class of indicator variables described in (3.4). Of course, sharing the same dog, i.e. aligning the dog nodes first, can provide strong evidence for aligning their owners, but this hierarchical structure will be left for the HEAT algorithm.

The constraints $C(\cdot)$ used to define the alignment set $\mathcal{K}$ are also derived from the attribute information and thus specific for each dataset. Continuing our dog park example, the user that is interested in aligning dogs can define the constraints so that the alignment set only contains dogs with the same breed as the one represented by $n_i$, while the user that is interested in aligning dogs' owners would constrain the alignment set to better reflect information directly related to an owner, e.g., owner's home address.

Given the inputs described above, the EAT algorithm (shown in Algorithm 1) loops through all potentially ambiguous nodes $n_i$ in graph $\mathcal{G}$ independently (line 2). For each of these nodes, it finds all nodes $n_k$ that satisfy the constraints for a specific $n_i$ to form an alignment set $\mathcal{K}_i$ (line 3). Then for each node in the alignment set, it counts the number of overlapping attributes between the "facts" associated with $n_i$ and those associated with $n_j$, along with any indicator variables (lines 4-12). Using the counts $c_k$ and the prior parameters $\alpha_k$, it calculates the posterior predictive value of alignment between a specific $n_i$ and $n_k$ (line 14). Finally, it returns a list of the expected probabilities between all nodes in $\mathcal{G}$ to all nodes in $\mathcal{G}'$.

**Algorithm 1** EAT
**Input:** $\mathcal{G}, \mathcal{G}', \boldsymbol{\alpha}, C(\cdot), A(\cdot)$
 1: Let $\mathbf{P} = \mathbf{0}$
 2: **for** $n_i \in \mathcal{G}$ **do**
 3: $\quad \mathcal{K}_i = \{n_k | n_k \in C(n_i)\}$
 4: $\quad$ **for** $n_k \in \mathcal{K}_i$ **do**
 5: $\quad\quad$ Let $c_k = 0$
 6: $\quad\quad$ **for** $n_j \in \mathcal{B}_i$ **do**
 7: $\quad\quad\quad$ Calculate $w_j$ from (3.3)
 8: $\quad\quad\quad$ **for** t from 1 to T **do**
 9: $\quad\quad\quad\quad$ Calculate $\iota_k$ from (3.4)
10: $\quad\quad\quad\quad$ $c_k = c_k + w_j[A(n_j) = A(\mathcal{B}'^t_k)]$
11: $\quad\quad\quad$ **end for**
12: $\quad\quad$ **end for**
13: $\quad\quad$ $c_k = \iota_k c_k$
14: $\quad\quad$ Calculate $p_{ik}$ from (3.6) and fill in $\mathbf{P}$
15: $\quad$ **end for**
16: **end for**
**Output: P**

The EAT algorithm calculates the probabilities of alignment between a specific type of node for a pair of graphs. However as mentioned in the dog park example above, aligning certain types of nodes, e.g., dogs, can improve the calculations for other types of nodes, e.g., dog owners. Thus, the HEAT algorithm wraps the EAT algorithm so that it can merge various types of nodes in a greedy fashion. In addition to the two graphs, it takes in a list of user defined "configurations" $[(C_1(\cdot), A_1(\cdot), \tau_1), \ldots]$, which contain a constraint attribute pair $C_1(\cdot), A_1(\cdot)$ for each "type" of node that the user wishes to align. Then in a hierarchical fashion it begins aligning nodes; starting with those for which a user is most confident that the dataset has evidence for. After each call of the EAT component, the HEAT algorithm merges the two graphs by merging nodes of the defined "type" who have a probability of alignment above a user defined threshold $\tau$. Then it moves on to the next configuration and aligns nodes in that "type". The pseudocode for the HEAT algorithm is show in Algorithm 2.

**Algorithm 2** HEAT
**Input:** $\mathcal{G}, \mathcal{G}', [(C_1(\cdot), A_1(\cdot), \tau_1), \ldots]$
 1: **while** $[(C_1(\cdot), A_1(\cdot), \tau_1), \ldots]$ **do**
 2: $\quad \mathbf{P} = $ Output of EAT
 3: $\quad$ **for** $n_i \in \mathcal{G}$ **do**
 4: $\quad\quad$ **if** $\max_k p_{ik} > \tau$ **then**
 5: $\quad\quad\quad$ k = arg $\max_k p_{ik}$
 6: $\quad\quad\quad$ Give $n_i$ in $\mathcal{G}$ same node id as $n_k$ in $\mathcal{G}'$
 7: $\quad\quad$ **end if**
 8: $\quad$ **end for**
 9: $\quad \mathcal{G}' = \mathcal{G} + \mathcal{G}'$ and $\mathcal{G} = \mathcal{G} + \mathcal{G}'$
10: **end while**
**Output:** Unified Graph

## 4 Experiments

In this section, we present the results of our algorithms on two real world problems. First we will review the data. Then we will discuss our choice of parameters. Finally we will show the performance of the HEAT algorithm on the dataset.

**4.1 Dataset** For the experiments in this section, we apply our HEAT algorithm to two datasets [11] curated by Pacific Northwest National Laboratory (PNNL), who served as the data providers and challenge problem evaluators for the MAA program.

The first dataset, which we will call the publication dataset, was generated from scientific bibliographic data from Elsevier's Scopus database [3]. PNNL's curated dataset consists of 500,000 different publications between 2018 and 2019 where each publication is associated with at least one research institution in New

York City. Along with research institutions, the publications also have associated titles, abstracts, keywords, and authors.

The second dataset, which we will call the transaction dataset, was generated from a Github datadump where over 7 million publicly available financial transactions were scraped from the Venmo public API [14]. These transactions took place between July - September 2018, October 2018, and January - February 2019 and contained the sender username, sender display name, receiver username, receiver display name, transaction message, and payment status.

Since one of the goals of the MAA program is to develop novel algorithms for aligning and merging graphs, PNNL created two graphs for each of the datasets where the graph associated with data before 2019 (pre2019 graph) is considered already "aligned" and the graph associated with data after 2019 (post2019 graph) is not "aligned". The nodes in the graph consist of the following:

- People, either as coauthors of a publication or the sender/receiver of a transaction, who have attributes in the form of names that have been obscured to be hashed values of the first initial and last name.

- Events, in the form of publications or transactions, who have attributes in the form of publication dates or transaction amounts respectively.

- Organizations, only in the publication graph, which have the organization's name and sporadically latitude and longitude attributes derived from the Wikidata knowledge base [20].

- Text, either from the keywords associated with the publications or the message required by Venmo for each transaction, which were mapped to their closest Qnodes in Wikidata.

The edges in the graph link all other nodes to the event nodes with edge attributes that described their relationship e.g. "Author" is an edge attribute for the edges between people and publication nodes. A subset of the publication graph is what is shown in Figure 1.

PNNL aligned the pre2019 graph by using the full first and last names of the people involved in the publications or transactions and mapping the raw text of the organization or text nodes to the closest Qnodes in Wikidata. For the post2019 graph, they also aligned the organization and text nodes, but left the people nodes unaligned. Thus, our challenge problem was to align the people nodes from the post2019 to the ones in the pre2019 graph where we observe noisy masked names of the people nodes in the form of a hash of the first initial and last name.

**4.2 Experiment Results** We applied the HEAT algorithm to the two datasets provided by PNNL which we described above. For both datasets, we extracted attributes in the form of the names of people and organizations and the keywords and messages of the publications and transactions respectively. While we are interested in aligning the people from the post2019 graph to the people in the pre2019 graph, we suspect that the other attributes on the graphs are also noisy. Thus, we first align the "text" nodes, keywords and messages, which PNNL mapped to Qnodes in Wikidata. Then, with less noisy "text" nodes, we align the people in the second iteration of the HEAT algorithm.

For ground truth, we received a file from PNNL containing the alignments of the people nodes from the post2019 to pre2019 graph. They created these alignments using the full first and last names in a similar fashion to how they created the aligned pre2019 graph. We did not receive this ground truth file after all algorithm development was complete; thus we developed our HEAT algorithm with only incremental validation in the form of performance curves received after each of the four rounds of independent evaluation.

In the first iteration, we set $C^1(\cdot)$ to constrain the alignment set to be "text" nodes whose text values, e.g. "machine learning", are with a normalized Levenshtein distance of 0.3. We merged all "text" nodes with a probability of alignment above $\tau = 0.7$. In the second iteration, we set $C^2(\cdot)$ so that the alignment set is only other people with the same hashed first initial plus last name. Specifically, for the publication dataset, because it is extremely unlikely for an author to change affiliations exactly between their last publication in 2018 and their first publication in 2019, we set the organization nodes to be an indicator variable.

In Figure 3 and Figure 4, we show the improved performance of the HEAT algorithm with the two iterations compared with the just using one iteration (EAT algorithm). The performance improvement on the publication dataset is due to having less noisy keywords, which improves the estimators for the probabilities of alignments. While the performance did not improve for the transaction dataset, we suspect this is because unlike keywords that describe a publication's area of research and thus are more likely to be reused by the same author, transaction messages do not hold the same significance.

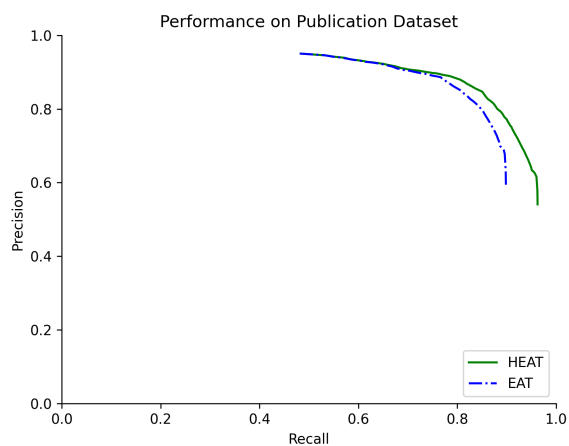

Figure 3: Precision and recall for one iteration (EAT) and two iterations (HEAT) on the publications dataset.

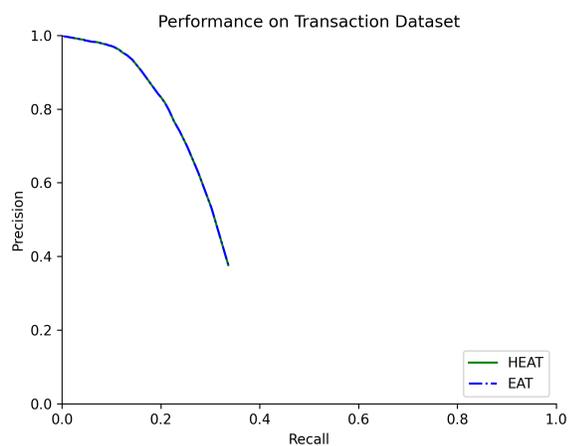

Figure 4: Precision and recall for one iteration (EAT) and two iterations (HEAT) on the transactions dataset.

Nevertheless, we show the noisy keywords and transaction messages that are aligned after the first iteration in the tables below. From Table 1, it is clear that almost all of the aligned keyword nodes represent similar areas of research. "Biomechanics" and "HIV type 1" are subclasses of "mechanics" and "HIV" respectively whereas "modulator" and "demodulator" and "ICD-10" and "ICD-11" (medical classification lists) are variations of the same topic. While "Adolescence" and "Adolescents" probably refer to a keyword for publications studying adolescents, the Qnodes refer to the painting by Salvador Dali and the American band respectively. The actual Qnode for an adolescent or teenager is Q1492760 making this mostly likely an example of where the keyword was incorrectly mapped to two separate Qnodes.

Table 1: Aligned Keywords For Publications

| node | keyword | node | keyword | prob. |
|---|---|---|---|---|
| Q41217 | "mechanics" | Q63202 | "Biomechanics" | 0.831 |
| Q1942300 | "modulator" | Q1185937 | "demodulator" | 0.833 |
| Q7175 | "mycology" | Q162555 | "oncology" | 0.842 |
| Q45127 | "ICD-10" | Q55695727 | "ICD-11" | 0.848 |
| Q15787 | "HIV" | Q18907320 | "HIV type 1" | 0.852 |
| Q2417154 | "Adolescence" | Q360484 | "Adolescents" | 0.903 |

Table 2 is a bit harder to interpret as the majority of messages are emojis. This is not surprising as Venmo auto-completes for emojis in their messages. The alignment of the Qnodes for "chicken" and "chick", which were not extracted into unicode emojis i.e. 🐔 and 🐣 or 🐥, but as literal strings, indicates that extraction of emojis can be extremely noisy leading to them being mapped to separate nodes. They can also be mapped to the incorrect Qnode as is the example of 🌊 which was mapped to the Qnode for a wind wave instead of the emoji-associated Qnode, Q87576769.

Table 2: Aligned Message Text For Transactions

| node | message | node | message | prob. |
|---|---|---|---|---|
| Q56683126 | 🎉 | Q50802362 | 💔 | 0.704 |
| Q165848 | 🌊 | Q56683126 | 🎉 | 0.744 |
| Q780 | "chicken" | Q1642639 | "chick" | 0.838 |

## 5 Discussion

We have proposed an algorithm for aligning entities based on their participation in similar events. Our model operates on a relatively unique scenario where events are the central component for constructing graphs. We showed good performance of our novel algorithm on two externally generated datasets where we did not have advanced access to ground truth.

Furthermore, because we have a generative model, it would be relatively easy to extend our approach to the problem of information completion. Just as we use the posterior predictive distribution of our Bayesian generative model to predict the probability of alignment from an ambiguous node to one in the "observed" graph, we can also our proposed model to "generate" the missing attributes of an ambiguous node based on its similarity to complete nodes in the "observed" graph.

**Acknowledgements**

This material is based upon work supported by the United States Air Force and DARPA under Contract No. FA8750-17-C-0156. Any opinions, findings and conclusions expressed in this material are those of the authors and do not necessarily reflect the views of the United States Air Force and DARPA.